\documentclass[twocolumn]{bmcart}
\usepackage{graphicx}
\usepackage{dcolumn}
\usepackage{bm}
\usepackage{sourceserifpro,sourcecodepro,sourcesanspro}
\usepackage[T1]{fontenc}
\usepackage{multirow}
\usepackage[version=3]{mhchem}
\usepackage[protrusion=true,expansion=true,final]{microtype}

\usepackage{hyperref}



\begin{document}
	\begin{frontmatter}
		\begin{fmbox}
			\dochead{Research}
			\title{MPFit: A robust method for fitting atomic resolution images with multiple Gaussian peaks}
			\author[
			addressref={aff1,aff2},
			email={mukherjeed@ornl.gov} 
			]{\inits{DM}\fnm{Debangshu} \snm{Mukherjee}}
			\author[
			addressref={aff1}
			]{\inits{LM}\fnm{Leixin} \snm{Miao}}
			\author[
			addressref={aff1}
			]{\inits{GS}\fnm{Greg} \snm{Stone}}
			\author[
			addressref={aff1},
			email={alem@matse.psu.edu} 
			]{\inits{NA}\fnm{Nasim} \snm{Alem}}

			\address[id=aff1]
			{
				\orgname{Department of Materials Science \& Engineering, The Pennsylvania State University},                    %
				\postcode{16802}                                
				\city{University Park},                              
				\cny{USA}                                    
			}
			\address[id=aff2]
			{
				\orgname{Center for Nanophase Materials Sciences, Oak Ridge National Laboratory},                    %
				\postcode{37831}                                
				\city{Oak Ridge},                              
				\cny{USA}                                    
			}
			
			\begin{abstractbox}
				
				\begin{abstract} 
					The standard technique for sub-pixel estimation of atom positions from atomic resolution scanning transmission electron microscopy images relies on fitting intensity maxima or minima with a two-dimensional Gaussian function. While this is a widespread method of measurement, it can be error prone in images with non-zero aberrations, strong intensity differences between adjacent atoms or in situations where the neighboring atom positions approach the resolution limit of the microscope. Here we demonstrate \textbf{mpfit}, an atom finding algorithm that iteratively calculates a series of overlapping two-dimensional Gaussian functions to fit the experimental dataset and then subsequently uses a subset of the calculated Gaussian functions to perform sub-pixel refinement of atom positions.  Based on both simulated and experimental datasets presented in this work, this approach gives lower errors when compared to the commonly used single Gaussian peak fitting approach and demonstrates increased robustness over a wider range of experimental conditions.
				\end{abstract}
				
				
			\end{abstractbox}
			
		\end{fmbox}
		
	\end{frontmatter}

\section*{\label{sec:Intro}Introduction}

The development of spherical aberration-correction for Scanning Transmission Electron Microscopy (STEM) imaging has been one of the biggest triumphs of electron microscopy over the past several decades, allowing the sub-\r{a}ngstr\"{o}m resolution imaging of crystal structures\cite{aberration1,aberration2,aberration3}. Several pioneering STEM experiments have demonstrated the feasibility of this technique for the direct visualization of atom positions from aberration-corrected STEM images and has proved itself an invaluable tool for sub-\r{a}ngstr\"{o}m resolution structural measurements\cite{SubAImaging,pico1,informationLimit,aberration4,pico2}. While the typical aberration-corrected STEM electron beam has a probe diameter approximately between 0.5\r{A} and 1\r{A}, supersampling scanning positions below the Nyquist-Shannon sampling limit and the subsequent fitting of the probe image with a two-dimensional Gaussian function allows the sub–pixel precision assignment of atom column positions from aberration corrected STEM datasets\cite{pico1,krivaneksub,pico3,pico4,pico5,pico6,pico7}. This technique has been used for quantitative atomic displacement measurements across thin films, 2D crystals, domain boundaries and has allowed the experimental observation of novel structural phenomena such as polar vortices\cite{lno_entropy,polarvortex,nelsonvortex,amin,gregstone,multivariate}.
 
While the Gaussian function fitting approach is an extraordinarily powerful technique, one noted shortcoming is that it assumes well separated atoms with no overlap, or negligible aberrations in the beam itself -- conditions that are only available under a certain limited set of imaging conditions\cite{polarvortex,nelsonvortex}. Typically, such an imaging setup uses a ring shaped annular detector with the outer and inner detector collection circles centered along the microscope optic axis. Such a configuration will have an inner collection angle of approximately 85--90 mrad to capture only the incoherently scattered electrons, and is conventionally referred to as High Angle Annular Dark Field STEM (HAADF-STEM) imaging\cite{pico1,williamscarter}. This mode of imaging is referred to as dark field imaging since atom columns themselves are bright due to electrons preferentially scattering from atomic nuclei as a consequence of Rutherford scattering from proton-electron Coulombic forces\cite{HAADF,HAADF_Si}. Since this Coulombic force experienced by the electron probe is directly proportional to the number of protons in the nucleus (Z), atom column images in HAADF-STEM datasets generate peaks with an almost linear relationship of intensity $\mathrm{\left( \propto Z^2 \right)}$ with the atomic number and is also referred to as Z contrast imaging\cite{Z_STEM,stemsim,quantitativeSTEM}.

Z-contrast imaging however is generally considered unsuitable for imaging lighter elements such as oxygen, boron or carbon\cite{gregstone,williamscarter,multivariate}. However, structural metrology for many scientifically important material systems such as ferroelectrics needs the imaging and quantification of lighter atoms as well as heavier elements\cite{spaldin_beginner,lines_glass}. This problem can be significantly mitigated in Bright Field STEM (BF-STEM) imaging, where rather than annular detectors a circular detector is used with the detector center coinciding with the optic axis of the microscope \cite{gregstone,abfstem}. The conventional collection angle ranges in BF-STEM imaging extend up to 15mrad, significantly lower than even the inner collection angle for HAADF-STEM\cite{abfstem}. Because unscattered electron beams are imaged by this technique, in contrast to HAADF-STEM vacuum is bright, while the atom positions have comparatively lower intensity. The ideal BF-STEM image would thus have an intensity profile complementary to the images obtained from HAADF-STEM imaging. However, in reality owing to the lower collection angles, atom positions are more blurred from aberrations that are more prominent in BF-STEM images\cite{BFvsHAADF}. Additionally, since BF-STEM images capture both light and heavy atom positions the inter-atomic distances are substantially smaller. These effects result in atom positions that are non-Gaussian in shape, and often have intensity overlaps and tails coming from their neighbors making position metrology challenging in BF-STEM images. 

\section*{\label{Methods}Methods}
\subsection*{\label{ssec:GaussFit}Fitting Atom Positions with Gaussians}

\begin{figure}
\centering
\includegraphics[width=\columnwidth]{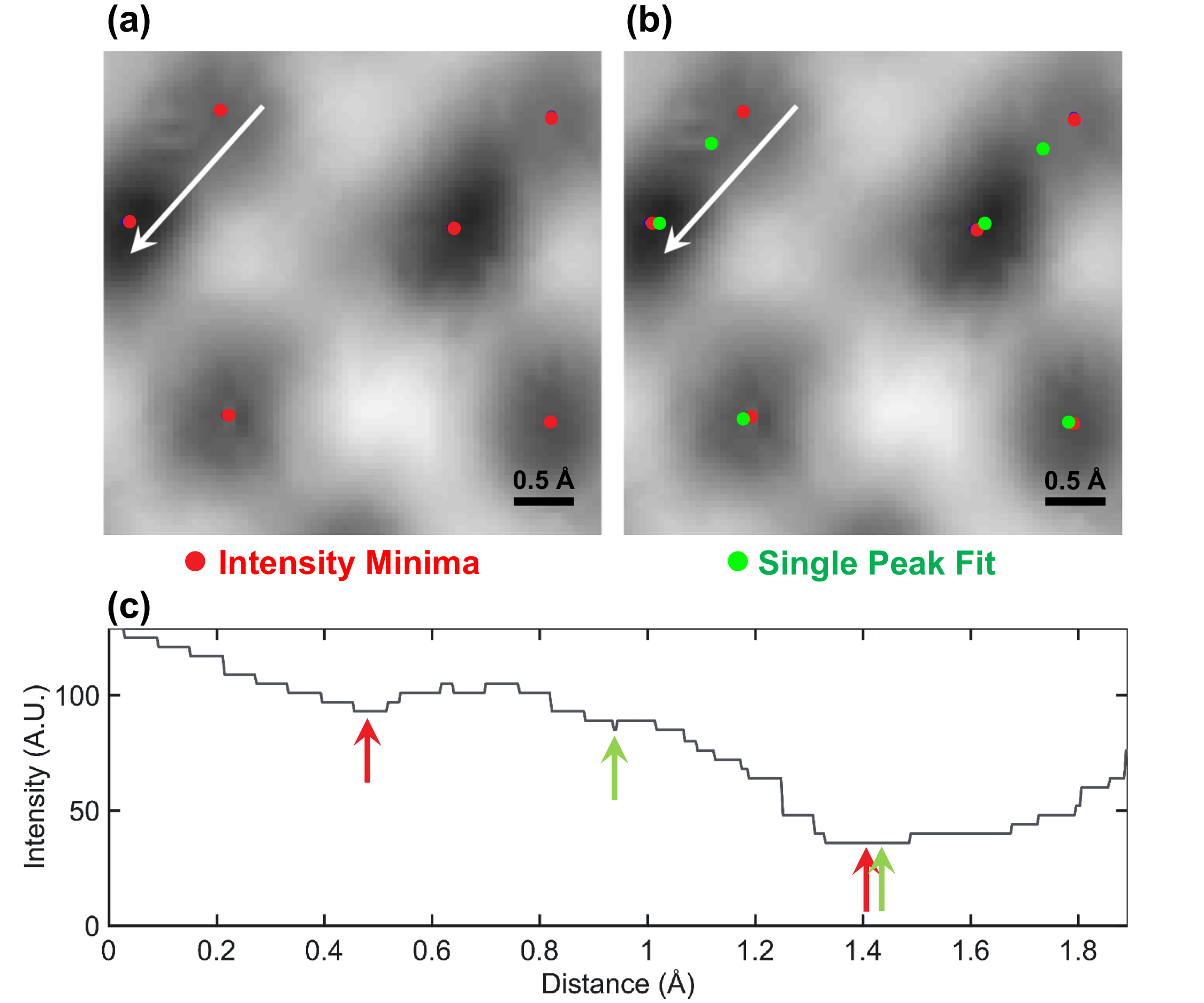}
\caption{\label{fig:SP_problem} \textbf{Error with single peak fitting on experimental data. (a)} BF-STEM image of \ce{LiNbO3} with the red dots referring to the intensity minima. \textbf{(b)} BF-STEM image shown in in \autoref{fig:SP_problem}(a) with the intensity minima and single peak fitting results overlaid in red and green respectively. \textbf{(c)} Intensity profile along the arrow shown in \autoref{fig:SP_problem}(a) and \autoref{fig:SP_problem}(b) with the red arrows referring to the intensity minima and the green arrows referring to the single peak fits.}
\end{figure}

The best modern aberration corrected microscopes can generate electron probes that are free of aberrations up to 30mrad, which corresponds to beam diameters that are of the order of 0.5 \AA, or 50pm at 200kV\cite{pico2,pico3}. Super-sampling the beam by a factor of five results in scan positions that are spaced approximately 10pm apart from each other. For HAADF-STEM images where oxygen atoms are not observed, inter-atomic distances from the low index zones are mostly of the order of 1.5 \AA, allowing enough distance between atoms so that they are well separated and thus an atom position can be reasonably approximated with a two-dimensional Gaussian intensity profile. Since the FWHM of this Gaussian is around 50-75pm, this allows the determination of the peak of the Gaussian intensity distribution with accuracies approaching 0.5pm\cite{pico7,nelsonvortex}. It is this combination of aberration-corrected imaging and Gaussian peak fitting that has enabled modern electron microscopy to reliably measure domain walls, grain boundaries, defects, and strain with single picometer precision, making STEM imaging so powerful. 

However, this approach runs into problems when applied to BF-STEM imaging. In \autoref{fig:SP_problem}(a), we show a typical BF-STEM image of \ce{LiNbO3} with 4.9 pm scanning pixel sizes. The darker regions in the image are the niobium and oxygen atom columns with the red dots corresponding to the intensity minima. While the intensity minima can be used as an initial estimate of atom positions, the error in such a measurement is at least of the order of the pixel size, which is 5pm in our case. This makes the error of measurement in BF-STEM an order of magnitude worse than the best HAADF-STEM results.  \autoref{fig:SP_problem}(b) demonstrates the same section of the BF-STEM image with the refined atom positions obtained from fitting the intensity distribution with a single Gaussian peak with green dots next to the intensity minima (red dots). A visual estimation shows that the fitted Gaussians do not reliably converge on the atom positions, and are often tens of picometers away when the intensity minima is weak, and the neighboring atom is close. In some cases, the refined atom position is in the middle of the two neighboring atom columns with no definite atomic intensity.

This can be quantitatively demonstrated by profiling the summed intensity distribution (\autoref{fig:SP_problem}(c)) from the region shown along the white arrows in \autoref{fig:SP_problem}(a) and \autoref{fig:SP_problem}(b). The red arrows in \autoref{fig:SP_problem}(c) correspond to the intensity minima, while the green arrows correspond to the Gaussian refined atom positions. The presence of an intense neighboring atom's intensity tail gives rise to a dip in the intensity away from the original minima, right in the middle of two atom columns and the Gaussian peak fitting technique converges to that local minima rather than the original position. Previous BF-STEM imaging has attempted in circumventing such issues by using a multi-parameter Gaussian peak, or performing image metrology through multivariate statistics rather than fitting each individual atoms\cite{gregstone,multivariate}. Both approaches require an initial knowledge of the crystal structure being imaged. Multi-parameter Gaussian fits need an estimation of the number and location of the nearest neighbors, and thus cannot be applied as a robust technique as it necessitates custom fitting equations for individual crystal structures.  In particular, this restriction limits the application of this method where more than one crystal orientation may be present. Here, we propose a novel multi-Gaussian refinement routine -- \textbf{mpfit} -- that does not require prior knowledge of the crystal structures being imaged and can robustly refine a wider variety of images by deconvolution of a subsection of the image into multiple overlapping two-dimensional Gaussians. Since HAADF-STEM image refinement requires less stringent conditions, our algorithm extends equally well to such systems too.

\subsection*{\label{ssec:MPFit}The \textbf{mpfit} algorithm}

\begin{figure*}[h!]
\centering
\includegraphics[width=\textwidth]{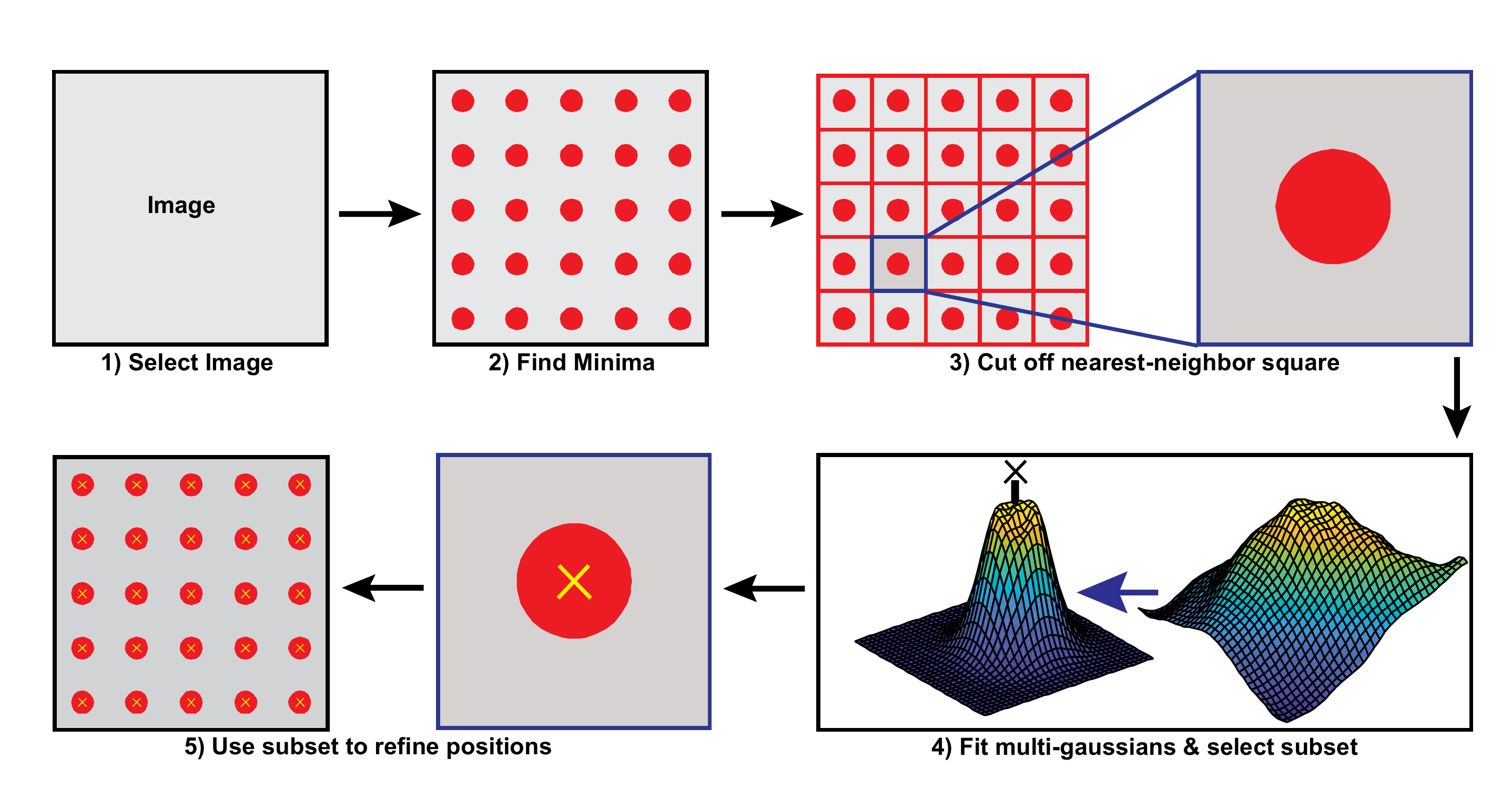}
\caption{\label{fig:schematic} \textbf{Schematic of the Procedure.} Red circles correspond to intensity minima or maxima for BF-STEM and ADF-STEM images respectively. The smaller squares surrounding the red dot refer to the nearest neighbor cutoff region while the yellow crosses refer to the refined atom positions.}
\end{figure*}

The Gaussian curve is a centrosymmetric curve with wide uses in single processing for approximating symmetric impulse functions\cite{multi_gaussian,gaussian_fitting}. Moreover, it has been demonstrated that given a sufficiently large number of Gaussians, any non-infinite signal can be approximated as a sum of overlapping Gaussians\cite{multi_gaussian,gaussian_fitting}. We use this insight and extend it into two dimensions by first modeling our observed atom intensity as a sum of overlapping Gaussians. The second key idea is to recognize that not all Gaussian functions that are approximating the region of interest are in fact originating from the atom whose position we are trying to refine. Thus the Gaussian functions are subsequently sorted and only a subset of them that approximate the atom position are used to refine the atom. The flow chart of our algorithm is illustrated in \autoref{fig:schematic}. The steps of the \textbf{mpfit} algorithm can be described as:
\begin{enumerate}
\item \textbf{Get intensity minima/maxima:} The initial starting point of this algorithm is the calculation of intensity maxima for inverted contrast BF-STEM images or ADF-STEM images. This can be implemented through standard MATLAB or Python peak finding routines. However in noisy images, sometimes a single atom may generate multiple maxima. To prevent this, if there are more than one maxima with the distance between the maxima smaller than the resolution of the microscope, the center of mass of this cluster of points is chosen as the starting reference point. 
\item \textbf{Calculate median inter-neighbor distance:} Following the identification of intensity minima, the median inter-peak distance is calculated, which is rounded to the nearest integer, which we call $\mathbf{\eta}$.
\item \textbf{Get region of interest:} The region of interest is a square with the intensity minima as the central pixel, with the sides of the square given by 
$\mathbf{s = 2\eta + 1}$, where \textbf{s} is the side of the square. Thus the $\mathbf{\left( \eta+ 1, \eta + 1 \right) }$ pixel in the square is the intensity minima that was the original starting point. Other regions of interest schemes, such as a Voronoi tesselation around the intensity minima actually demonstrate comparatively worse results (see \autoref{fig:voronoi_ROI} in the supplemental material). 
\item \textbf{Fit iteratively with Gaussians:} The region of interest is then fit by a single 2D Gaussian function with a user determined tolerance factor. The tolerance factor refers to the mean absolute difference in intensity between the fitted gaussian and the original data. The fitted Gaussian function is then subtracted from the original region of interest, and the residual is subsequently fitted again. This process continues for a pre-determined number of iterations, with the sum of all the Gaussians then subsequently representing the original region of interest. In the authors' experience, the tolerance factor is less important than the number of iterative Gaussians used, with reasonable accuracy and speed being obtained with a tolerance of $\mathrm{10^{-12}}$ and 12 to 16 iterations for the \textbf{mpfit} examples presented in this text. We deal with the background intensity by normalizing each ROI cell from 0 to 1 before starting the estimation of the individual Gaussians.
\item \textbf{Sort peaks and get the refined position:} The Gaussian peaks are then subsequently sorted based on their distance from the original minima (step 1) with only those  peak positions whose distances are less than $\mathbf{\frac{\eta}{2}}$ from the minima/maxima (step 1) used for refinement. The refined atom position is then the weighted average peak position of all the Gaussians that lie within this selected region, with the peak amplitudes being the weights used.
\end{enumerate}

\section*{\label{sec:Results}Results and Discussions}
\subsection*{\label{ssec:Simulated}Results on Simulated BF-STEM Images}

\begin{figure*}[h!]
\centering
\includegraphics[width=\textwidth]{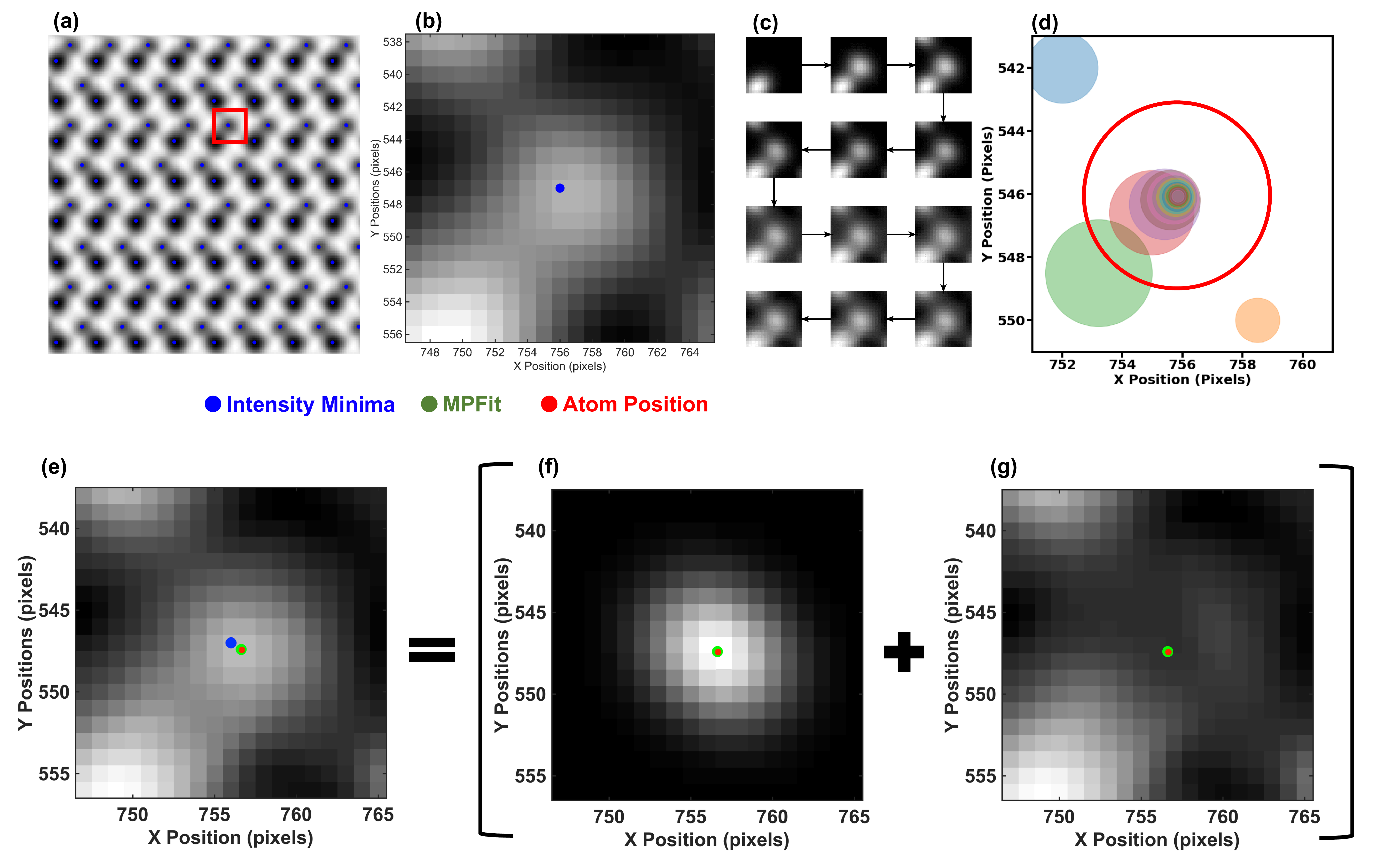}
\caption{\label{fig:SIMpeakEvolve} \textbf{Evolution of Gaussian peaks for \emph{simulated} data. (a)} Simulated BF-STEM image of \ce{LiNbO3} with the intensity minima overlaid as blue dots. \textbf{(b)}  Calculation region of interest, demonstrated as the red box in \autoref{fig:SIMpeakEvolve}(a) of the simulated BF-STEM image with the intensity reversed, with the blue spot corresponding to the intensity minima. \textbf{(c)} Evolution of the sum of the Gaussian peaks over multiple iterations. \textbf{(d)} Contributions of the Gaussian peaks scaled to their amplitudes with larger spheres corresponding to peaks with higher amplitudes. The red circle refers to the region from which the Gaussian peaks were selected from. \textbf{(e)} Equivalent summation of multiple Gaussian peaks with the blue point corresponding to the intensity minima, the green point corresponding to the atom position calculated by the \textbf{mpfit} algorithm and the red point corresponding to the atom position. \textbf{(f)}  Contribution from the atom whose positions is being measured. \textbf{(g)} Contribution from nearest neighbors.}
\end{figure*}

The efficiency and accuracy of the \textbf{mpfit} algorithm was tested on simulated BF-STEM images of \ce{LiNbO3}. The advantage of simulated data is that the accurate atom positions are already known and can be compared with \textbf{mpfit} results. This allows the estimation of the relative errors of the single Gaussian and the multiple Gaussian \textbf{mpfit} approaches, with the simulation parameters outlined in \autoref{tab:Conditions}. Following the steps of the algorithm outlined in \autoref{fig:schematic}, the intensity minima were first calculated for the simulated image, with \autoref{fig:SIMpeakEvolve}(a) demonstrating the simulated BF-STEM image of \ce{LiNbO3} with the intensity minima overlaid as blue dots. These intensity minima are subsequently used to calculate the median nearest neighboring distance $\mathbf{\left( \eta \right)}$ between the minima.  Based on the calculated $\mathbf{\eta}$ value, the region of interest for this image is demonstrated for one of the atoms as a red square in \autoref{fig:SIMpeakEvolve}(a). The region of interest for that atom is shown in \autoref{fig:SIMpeakEvolve}(b) with the contrast inverted and the intensity minima for the atom in question overlaid as a blue dot. As could be ascertained from \autoref{fig:SIMpeakEvolve}(b), the intensity distribution from the bottom left atom partially overlaps with the atom position we are aiming to refine, precisely indicating the scenario where single peak Gaussian fitting approaches often give erroneous results.
 
Sixteen iteration steps were chosen to represent this section of the image, as per step four of the algorithm. The first twelve of these iteration steps and the evolution of the Gaussian summation is shown in \autoref{fig:SIMpeakEvolve}(c). The calculation of the Gaussian is performed by taking in the entire image, and calculating a two-dimensional Gaussian peak with the smallest absolute difference with the initial region of interest. Multiple different Gaussian fitting approaches can be used, with the fitting equation used in this approach expanded in \autoref{eq:Gauss}. As could be observed from \autoref{fig:SIMpeakEvolve}(c) the summation of the Gaussian peaks starts to approximate the region of interest within only a few iterations. This demonstrates that the iteration number chosen was sufficient enough to capture the complexities of the data being fitted. It is even more interesting to look at the result of the first iteration, which is mathematically equivalent to the single Gaussian peak fitting approach. As the first iteration in \autoref{fig:SIMpeakEvolve}(c) shows, the single Gaussian peak fitting approach is a special case of the \textbf{mpfit} algorithm, where the number of iterations is one. According to this image, the first Gaussian peak does not exist near the center of the image, and is extracted towards the bottom left corner.  The central peak related to the atomic column of interest is captured in the second iteration rather than the first, thus visually demonstrating why the single Gaussian approach fails in some cases. 

While it may be possible to adjust the calculation of the region of interest to capture the atom position accurately, this approach necessitates tinkering with multiple different collection areas and a non-uniform solution for all the atoms in the image. \textbf{mpfit} on the other hand, removes the necessity for such complicated user modifications, allowing the estimation of all the Gaussian peaks that contribute to the final image. The individual peak positions are  visually represented as a function of the iteration number in \autoref{fig:SIMpeakEvolve}(d) where the radius of the spheres are scaled to the amplitude of the Gaussian calculated -- with the X and Y position of the sphere referring to the location  of the Gaussian peak. Only Gaussian peaks lying below a certain distance from the intensity minima (shown as the red circle in \autoref{fig:SIMpeakEvolve}(d)) are used for the estimation of the refined position. As could be observed, the peak obtained from the the most intense Gaussian in green is actually assigned to the neighboring atom, and two other peaks are also assigned to two neighboring atoms, and only the subset of Gaussian peaks lying within the red circle are used for refinement. Summing all the Gaussian functions together we obtain \autoref{fig:SIMpeakEvolve}(e), which shows close fidelity to the input data (\autoref{fig:SIMpeakEvolve}(b)).

Based on the final step of the algorithm, the Gaussian peaks are assigned either to neighboring atoms or the central atom depending upon the distance of the peak center from the initial intensity minima.  As can be seen in \autoref{fig:SIMpeakEvolve}(e), the intensity minima is not always a reliable estimator of the actual atom position, but the \textbf{mpfit} algorithm converges extraordinarily close to the actual atom position (green and red dots overlapping), demonstrating its superiority. The representative summation of the Gaussian summation can thus be broken down into two components – the Gaussian peaks that were used for atom position refinement, the sum of which is visualized in \autoref{fig:SIMpeakEvolve}(f) and the Gaussian peaks that were further off, and assigned as contributions of neighboring atom intensity tails – represented in \autoref{fig:SIMpeakEvolve}(g). Thus, the combination of the main atom and the neighboring contributions gives rise to the total intensity profile that was observed.

\begin{figure*}[h!]
\centering
\includegraphics[width=\textwidth]{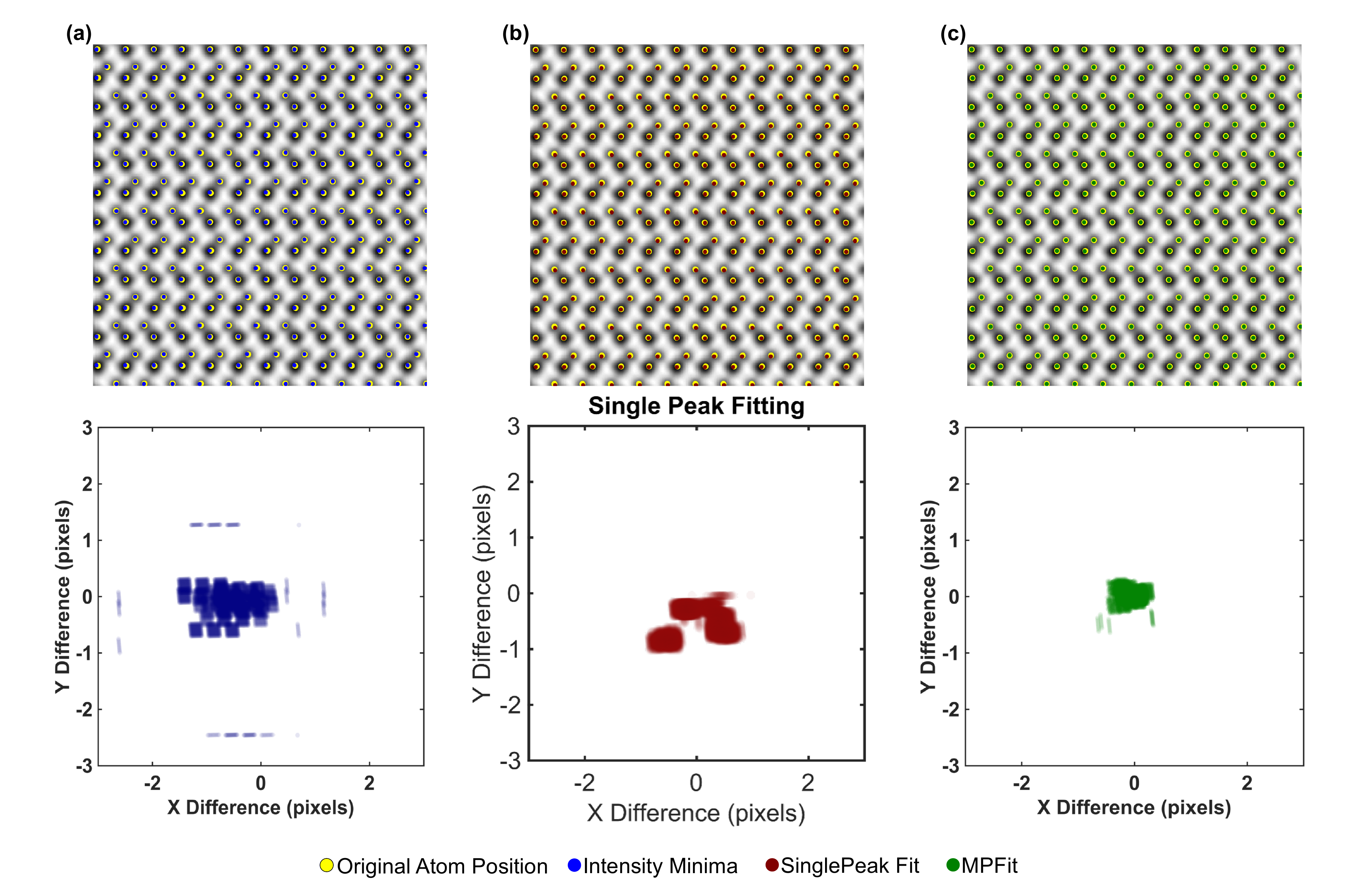}
\caption{\label{fig:SIMcalc} \textbf{Calculated positions. (a)} Simulated \ce{LiNbO3} BF-STEM data with the original atom positions and the intensity minima overlaid in yellow and teal respectively. \textbf{(b)} Simulated \ce{LiNbO3} BF-STEM data with the original atom positions and the atom positions obtained by fitting a single Gaussian peak overlaid in yellow and red respectively. \textbf{(c)} Simulated \ce{LiNbO3} BF-STEM data with the original atom positions and the atom positions calculated via the \textbf{mpfit} algorithm overlaid in yellow and green respectively.}
\end{figure*}

We further evaluated the accuracy of the \textbf{mpfit} algorithm for an entire image rather than a single atom. \autoref{fig:SIMcalc} shows and compares the three different atom position metrology techniques – intensity minima/maxima, single Gaussian peak fitting and the \textbf{mpfit} algorithm with each other respectively. \autoref{fig:SIMcalc}(a) demonstrates the intensity minima itself may not be coincident with the ideal atom positions due to minute intensity variations that are not accurately captured given a limited detector dynamic range, with the errors of the order of a single pixel. As a result, the intensity minima atom positions are clustered at several different clustering values, which can be understood based on the fact that results from the intensity minima are always on the order of a pixel. Thus compared to position refinement algorithms, just the minima itself is incapable of sub-pixel precision metrology. \autoref{fig:SIMcalc}(b) demonstrates the difference of the single peak approach from the ideal atom positions, with the results being clustered into three distinct clusters. This can be understood based on the fact that there are three separate types of intensity distributions in the simulated data. For well-separated atoms, the single peak and the atom positions show close agreement, which generates the central cluster. However, there are also atom columns, where the neighboring atoms are either on the top left or the top right, giving rise to the two extra clusters -- demonstrating the shortcomings of this approach when the intensity distributions of neighboring atoms approach the resolution limit of the electron microscope. The results from the \textbf{mpfit} algorithm, demonstrated in \autoref{fig:SIMcalc}(c), are on the other hand clustered in a region less than 0.5pm across from the known atom positions -- demonstrating it's accuracy. However in the authors' experience, the \textbf{mpfit} technique fails to converge for the edge atoms, which shows up as atom positions that are not clustered and have a higher error. For the rest of the atoms in the image, however, \textbf{mpfit} is significantly superior to the other approaches. 

\subsection*{\label{ssec:Experimental}Results on Experimental BF-STEM Images}

\begin{figure*}[h!]
	\centering
	\includegraphics[width=\textwidth]{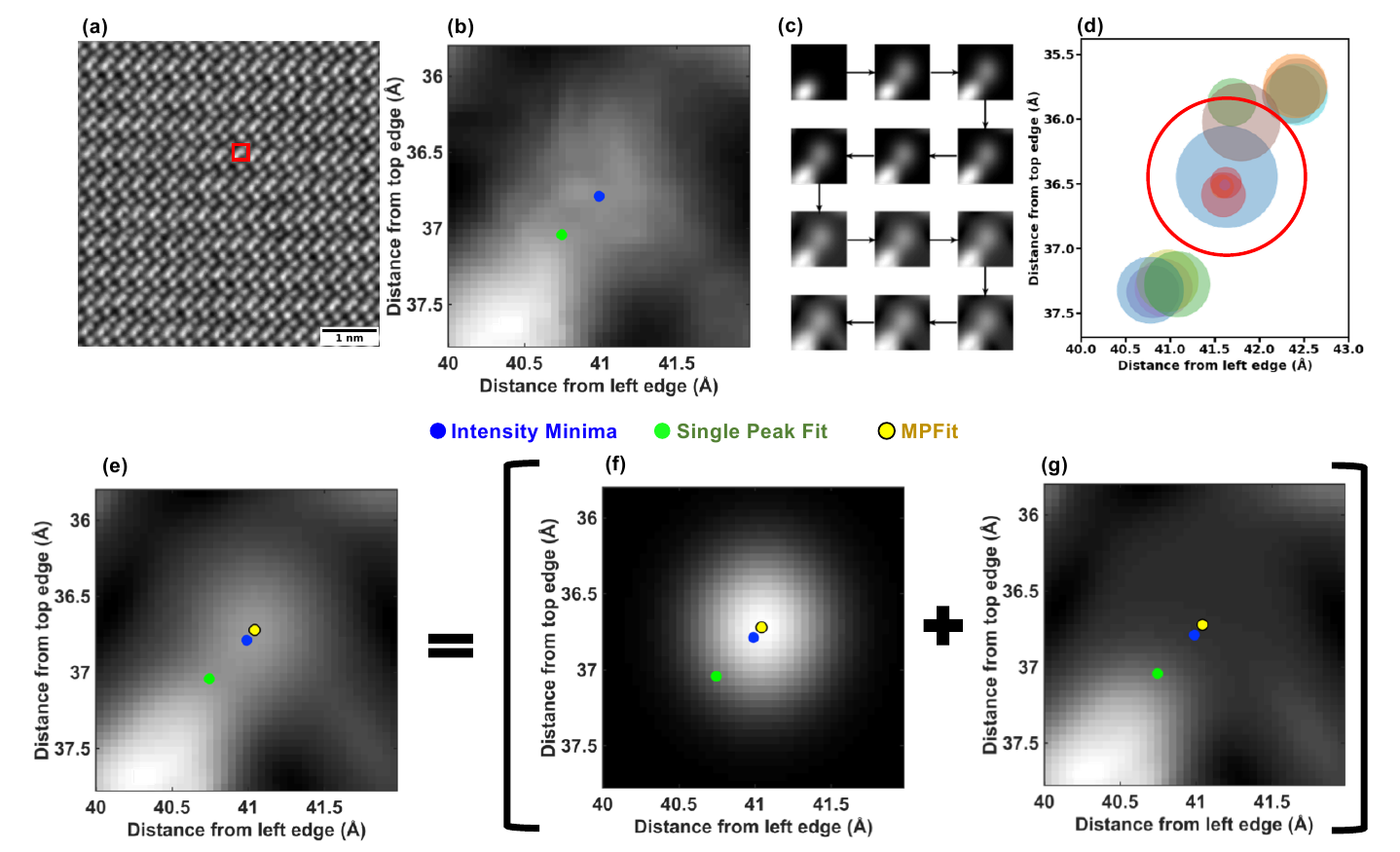}
	\caption{\label{fig:EXPpeakEvolve} \textbf{Evolution of Gaussian peaks for \emph{experimental} data. (a)} Experimental inverted contrast BF-STEM image of \ce{LiNbO3}. \textbf{(b)} Calculation region of interest of an experimental BF-STEM image with the intensity reversed from the region marked by the red box in \autoref{fig:EXPpeakEvolve}(a), with the blue point corresponding to the intensity minima, and the green point corresponding to the position calculated by fitting a single Gaussian peak.  \textbf{(c)} Evolution of the sum of the Gaussian peaks over multiple iterations. \textbf{(d)} Contributions of the Gaussian peaks scaled as a function of their amplitude. Peaks lying outside the red circle are assigned to neighboring atoms. \textbf{(e)} Equivalent summation of multiple Gaussian peaks with the blue point representing the location of the intensity minima, the green point the position calculated by fitting a single Gaussian peak and the yellow point representing the atom position calculated by the \textbf{mpfit} algorithm. \textbf{(f)} Contribution from the atom whose positions is being measured. \textbf{(g)} Contribution from nearest neighbors.}
\end{figure*}

Along with simulated datasets, we additionally performed position metrology on experimental BF-STEM images of \ce{LiNbO3} viewed from the $\mathrm{\left[ 1\bar{1}00 \right] }$ zone axis\cite{lno_entropy}. The results were obtained through STEM imaging in a spherical aberration corrected FEI $\mathrm{Titan^3}$ transmission electron microscope, corrected for upto third order spherical aberrations. Imaging was performed at a camera length of 145mm and the BF-STEM images were collected using Gatan detectors with an outer collection semi-angle of 15mrad, using scanning pixel step sizes of 9.8pm. 

In contrast to simulated datasets, the exact ideal atom positions are not known owing to specimen drift, thermal vibrations, signal to noise ratio, and localized imperfections in the crystal lattice. \autoref{fig:EXPpeakEvolve}(a) demonstrates a region of interest in an experimental dataset, with the intensity reversed, with \autoref{fig:EXPpeakEvolve}(b) showing a section of the image marked by the red box in \autoref{fig:EXPpeakEvolve}(a). The blue dot in \autoref{fig:EXPpeakEvolve}(b) corresponds to the intensity minima, while the green dot represents the position calculated by the single Gaussian peak fitting approach. As can be visually ascertained, the calculated atom position does not correspond to the atom position, and thus is an inaccurate representation. Following step 4 of the algorithm, and similar to the procedure outlined in \autoref{fig:SIMpeakEvolve}(c), the region of interest is represented by a succession of closely spaced two-dimensional Gaussian peaks over sixteen iteration steps,with the contribution from the first 12 steps shown in \autoref{fig:EXPpeakEvolve}(c). The individual Gaussian peaks that contribute to the final representation of the region of interest are pictorially represented in \autoref{fig:EXPpeakEvolve}(d), with the radius of the circle corresponding to the amplitude of the Gaussian. Peaks that are further from the original intensity minima by more than twice the median inter-peak distance (which is indicated by the red circle) are assigned to the neighboring atoms, and only peaks lying inside the red circle are used to calculate the refined atom position. 

\autoref{fig:EXPpeakEvolve}(b) demonstrates the initial experimental data, while \autoref{fig:EXPpeakEvolve}(e) demonstrates the final summation from the sixteen Gaussian peaks with visual inspection revealing close correspondence between the experimental and represented data. Extending the number of iterations would allow progressively smaller Gaussian peaks resulting in better correspondence, but would also increase the demand for computational resources without a correspondingly significant increase in precision. The intensity minima are overlaid on the images in blue, with the results from the single peak fit approach in green and the \textbf{mpfit} results in yellow respectively. Thus the \textbf{mpfit} algorithm accurately determines the atom location rather than converging to saddle points created from intensity tails from neighboring atoms. \autoref{fig:EXPpeakEvolve}(f) represents the sum of the Gaussians that represents the atom being refined and \autoref{fig:EXPpeakEvolve}(g) represents the contribution from the intensity tails from the neighboring atoms, and is calculated from the Gaussian peaks represented with red borders in \autoref{fig:EXPpeakEvolve}(d).

\begin{figure*}[h!]
\centering
\includegraphics[width=\textwidth]{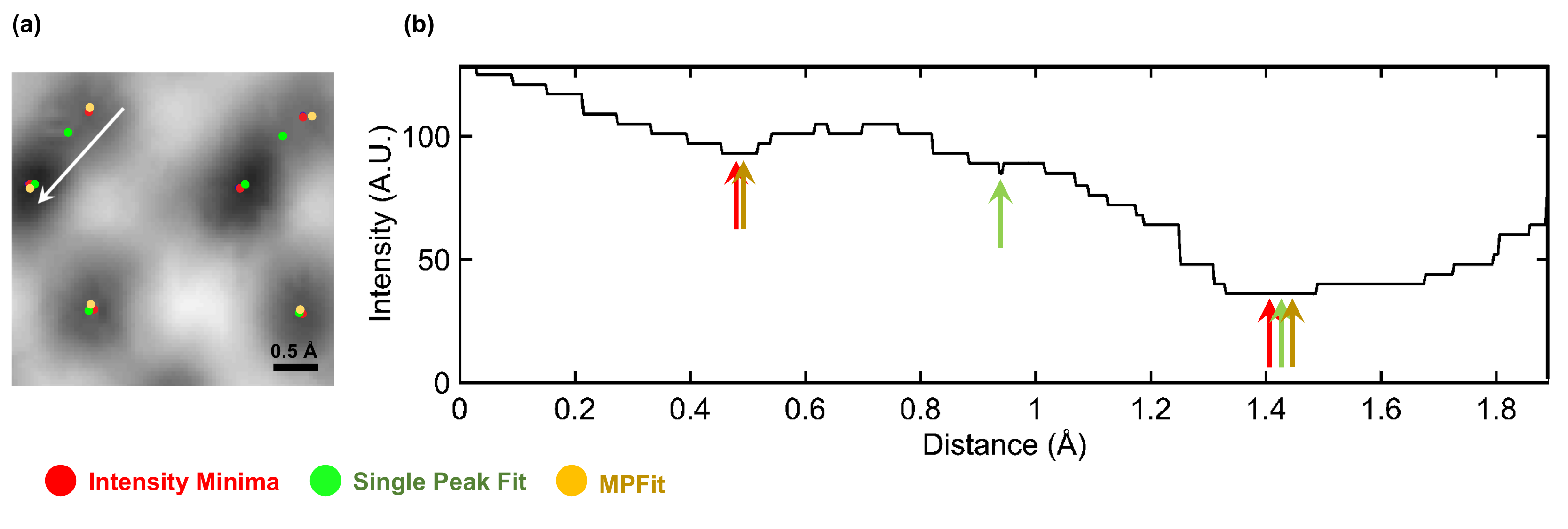}
\caption{\label{fig:EXPcompare} \textbf{Comparing \textbf{mpfit} with single peak fitting on experimental data. (a)} Experimental \ce{LiNbO3} BF-STEM data with the red points referring to the intensity minima, green points referring to the fitted positions as calculated by the single peak approach, and yellow points being the points as calculated by the \textbf{mpfit} approach. \textbf{(b)} Intensity profile of the image along the white arrow, with the red arrows corresponding to intensity minima, green arrows to single peak refinement results, and yellow arrows referring to the \textbf{mpfit} refinement results.}
\end{figure*}

Returning back to the original experimental BF-STEM image in \autoref{fig:SP_problem}(a), we revisit that experimental data in \autoref{fig:EXPcompare}(a), comparing the results obtained with the \textbf{mpfit} approach. As could be visually ascertained, while the single peak fit approach fails in some of the cases, the \textbf{mpfit} approach reliably refines to the atom position, which can also be ascertained by the intensity profile demonstrated in \autoref{fig:EXPcompare}(b). 

\subsection*{\label{ssec:Comparisons}Comparisons with other algorithms}

\begin{figure*}[h!]
	\centering
	\includegraphics[width=\textwidth]{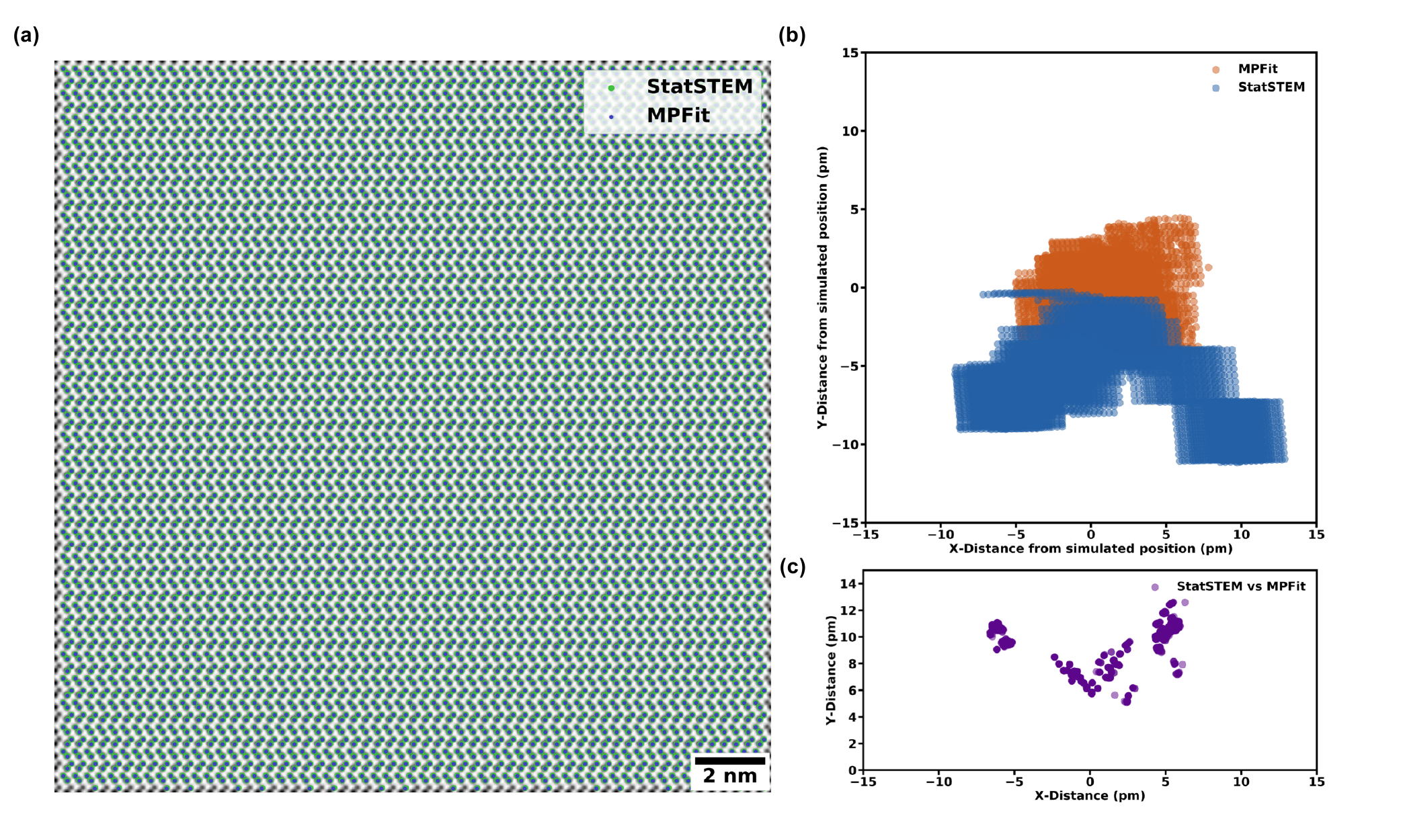}
	\caption{\label{fig:statstem} \textbf{Comparision of \textbf{mpfit} and \textbf{StatSTEM} on simulated images. (a)} Overlaid results from \textbf{StatSTEM} and \textbf{mpfit} on a simulated \ce{LiNbO3} dataset. \textbf{(b)} Distance from known atom positions and the calculated positions from \textbf{StatSTEM} (in blue) and \textbf{mpfit} (in orange).  \textbf{(c)} Distance between \textbf{StatSTEM} and \textbf{mpfit} results in picometers.}
\end{figure*}

Several other specialized algorithms have been designed to quantify atom positions in electron microscope datasets, such as \textbf{Atomap}\cite{atomap}, \textbf{StatSTEM}\cite{statstem} and \textbf{oxygen octahedra picker}\cite{OOpicker}. The \textbf{Atomap} algorithm uses principal component analysis to obtain denoised STEM images and finds the center of mass based on the initial guess of local intensity maxima or minima. Using the center of mass as the starting estimate, it then subsequently approximates a two-dimensional gaussian to locate the estimated position of atoms. \textbf{Atomap} can additionally sort the different species of atom columns in the image and analyze them individually. \textbf{StatSTEM} on the other hand is a model-based fitting algorithm for extraction of the atom position information from STEM images. \textbf{StatSTEM} models the atoms in the images as the superposition of two-dimensional gaussian peaks, and since this is a model based technique it requires prior knowledge of the crystal structure of the sample being imaged to give a better estimation of the initial guess. After obtaining the initial guess, the algorithm will go through iterations to reach the least squares estimation of fitting parameters, and then determines the position. \textbf{oxygen octahedra picker} is a software specialized in identifying the octahedra rotations in the \ce{ABO3} perovskite oxides. It sorts out the oxygen and B atom positions and provides users the option of selecting a fast center of mass estimation or a slower peak fitting with two-dimensional gaussians. It exhibits an impressive accuracy of as small as 3 pm in simulated HAADF images. However, the existing methods still possess limitations in practical cases -- with neither the \textbf{oxygen octahedra picker} and the \textbf{Atomap} software being able to process STEM images where atomic columns being measured will have intensity contributions from their neighbors. Thus both these approaches work well for well-separated atom columns in HAADF images, but face accuracy penalties with BF-STEM images. While \textbf{StatSTEM}'s model based algorithm is able to solve the overlapping issue by assuming atom columns as overlapping 2D Gaussian peaks, its iterative model fitting process is computationally intensive, and requires prior knowledge of the crystal structure being imaged. As demonstrated in \autoref{fig:statstem}(a), visually there is almost no difference between the fitting results of \textbf{StatSTEM} versus \textbf{mpfit}, with \textbf{StatSTEM}'s results being slightly off-centered from \textbf{mpfit}'s estimation. Comparison of the results in \autoref{fig:statstem}(b) demonstrates that both technique give results that are less than a pixel apart from each other, with \textbf{mpfit} outperforming \textbf{StatSTEM}. The standard deviation $\mathrm{\left( \sigma \right)}$ of \textbf{mpfit}'s estimation from known atom positions is 1.49pm compared to a $\mathrm{\sigma}$ of 3.31pm from \textbf{StatSTEM}.

\section*{\label{sec:Conclusions}Conclusions}

While it may be possible to assume from the results presented here that the single Gaussian peak fitting approach fails to converge to atom solutions and gives erroneous results, it actually performs perfectly adequately for the majority of STEM experiments. However, for certain non-ideal imaging conditions, the single Gaussian peak fitting approach fails, while \textbf{mpfit} accurately obtains precise atom positions. For well-separated atoms, the results from \textbf{mpfit} and a single Gaussian refinement are in fact identical. Additionally, it has to be kept in mind, that even with parallelization implemented, the \textbf{mpfit} algorithm solves for over ten Gaussian peaks in a batch process. On the other hand, the single Gaussian approach solves for just one peak, thus making the single Gaussian approach faster by at around an order of magnitude. 

Future planned improvements include solving for neighboring peaks simultaneously using the tail functions to deconvolve the full obtained image as an independent set of impulse functions originating from individual atoms. Additionally, atom columns whose separation distances are below the resolution limit of the microscope may be particularly suited for this approach, by the deconvolution of the observed impulse function into two closely separated Gaussians and enabling the super-resolution metrology of atom positions from STEM datasets.

Thus, our results demonstrate that the \textbf{mpfit} algorithm can reliably and robustly refine the sub-pixel precision of atoms even without \emph{a priori} knowledge of the underlying crystal structure. Additionally, since the single Gaussian approach is a special case of the \textbf{mpfit} approach with the total number of iterations as one, this approach will also work for ADF-STEM images, enabling a single approach to the metrology of a wide variety of STEM data. The results are superior to existing algorithms, and exceeds the state of the art -- \textbf{StatSTEM} in accuracy, with the added advantage of being agnostic to the crystal structure being imaged.

\begin{backmatter}

\section*{List of Abbreviations}
STEM, ADF, BF, HAADF
\section*{Availability of data and materials}
\textbf{mpfit} code in MATLAB is freely available from the \href{https://github.com/dxm447/mpfit}{MPfit Github repository}.  The Python codes are available \href{https://github.com/dxm447/stemtools/blob/no_ptycho/stemtools/afit/atom_positions.py}{in the \texttt{afit} module of \texttt{stemtools}}. The datasets used and/or analysed during the current study are available from the corresponding author (D.M.) on reasonable request.
\section*{Competing interests}
The authors declare that they have no competing interests.
\section*{Funding}
The authors acknowledge funding support from the Penn State Center of Nanoscale Science, an NSF MRSEC, funded under the grant number DMR-1420620.
\section*{Authors' Contributions}
D.M. and G.S., advised by N.A. designed the algorithm. D.M. developed the MATLAB and Python subroutines for the algorithm implementation. D.M. and L.M., advised by N.A. analyzed the data and developed the algorithm. D.M. performed the electron microscopy simulations. D.M. wrote the manuscript. N.A. edited the manuscript. All authors commented on the final manuscript.
\section*{Acknowledgements}
The authors would like to acknowledge Dr. Jason Lapano and Prof. Venkatraman Gopalan of Penn State for helpful discussions on atom fitting metrology.

\section*{Supplemental Information}

\subsection*{\label{ssec:gausscalc}Gaussian calculation parameters}

The Gaussian peaks were calculated based on \autoref{eq:Gauss}
\begin{multline}\label{eq:Gauss}
\mathrm{Z\left( x,y \right) = Ae^{\frac{\left( \left( \left( x-x_0 \right) \cos\theta \right) + \left( \left( y-y_0 \right) \sin\theta \right) \right)^2}{\sigma_x}} \times} \\
	                            \mathrm{e^{\frac{\left( \left( \left( x-x_0 \right) \sin\theta \right) + \left( \left( y-y_0 \right) \cos\theta \right) \right)^2}{\sigma_y}}}
\end{multline}
where $Z\left( x,y \right) $ is the Gaussian output as a function of $x$ and $y$, $\sigma_x$ and $\sigma_y$ are the two normal distributions in the $x$ and $y$ directions, $x_0$ and $y_0$ are the position of the Gaussian peak, $A$ is the amplitude of the Gaussian peak and $\theta$ is the rotation in the counter-clockwise direction of the two-dimensional Gaussian peak. 

Thus given a set of $x$,$y$, and $z$ values from the experimental region of interest, a Gaussian curve is estimated from \autoref{eq:Gauss} such that:
\begin{equation}
\sum_{x,y}\left( z - Z \right) = \tau
\end{equation}
where $\tau$ is the tolerance, which was $\mathrm{10^{-8}}$ for our implementation. 

The Equation itself is calculated through the least-squares approach using the trust region reflective algorithm. Trust-region algorithms are an evolution of Levenberg-Marquardt (LM) algorithms. However, compared to the LM algorithms, this algorithm is curvature independent and is thus computationally significantly faster\cite{levenberg,marquardt,trust_region}.

\subsection*{\label{ssec:sim_param}Simulation Parameters}

The \ce{LiNbO3} images were simulated using the MacTempasX software, with the simulation  parameters enumerated in \autoref{tab:Conditions}\cite{mactempasx}.

\begin{table}[h!]
	\caption{\label{tab:Conditions}BF-STEM simulation conditions in MacTempasX}
	\begin{tabular}{|c|c|}
	\hline
	\textbf{Experimental Condition}	& \textbf{Value}\\
	\hline
	\hline
	Crystal Structure	& \ce{LiNbO3} \\
	\hline
	\multirow{3}{*}{Debye-Waller Parameters} & $\mathrm{u_{Li} = 0.67}${\AA} \\
	 & $\mathrm{u_{Nb} = 0.3924}${\AA} \\
	 & $\mathrm{u_O = 0.5}${\AA}\cite{LNO_XRD} \\ 
	\hline
	\multirow{3}{*}{Lattice Parameters} & a = 5.172{\AA} \\
	 & b = 5.172{\AA} \\ 
	 & c = 13.867{\AA}\cite{LNO_neutron} \\
	\hline
	Space Group & 161 (R3c)\cite{megawLNO}\\
	\hline
	Zone Axis	& $\mathrm{\left[ 1 \bar{1} 00 \right] }$\\
	\hline
	Accelerating Voltage	& 200kV\\
	\hline
	Inner Collection Angle	& 0mrad\\
	\hline
	Outer Collection Angle	& 15mrad\\
	\hline
	Cells & $\mathrm{1 \times 5}$\\
	\hline
	Frozen Phonons & 10\\
	\hline
	Slices per Unit Cell & 5\\
	\hline
	Probe Semi-Angle	& 28mrad\\
	\hline
	\end{tabular}
\end{table}

\subsection*{\label{ssec:voronoi}Voronoi vs square ROI}
\begin{figure*}
	\centering
	\includegraphics[width=\textwidth]{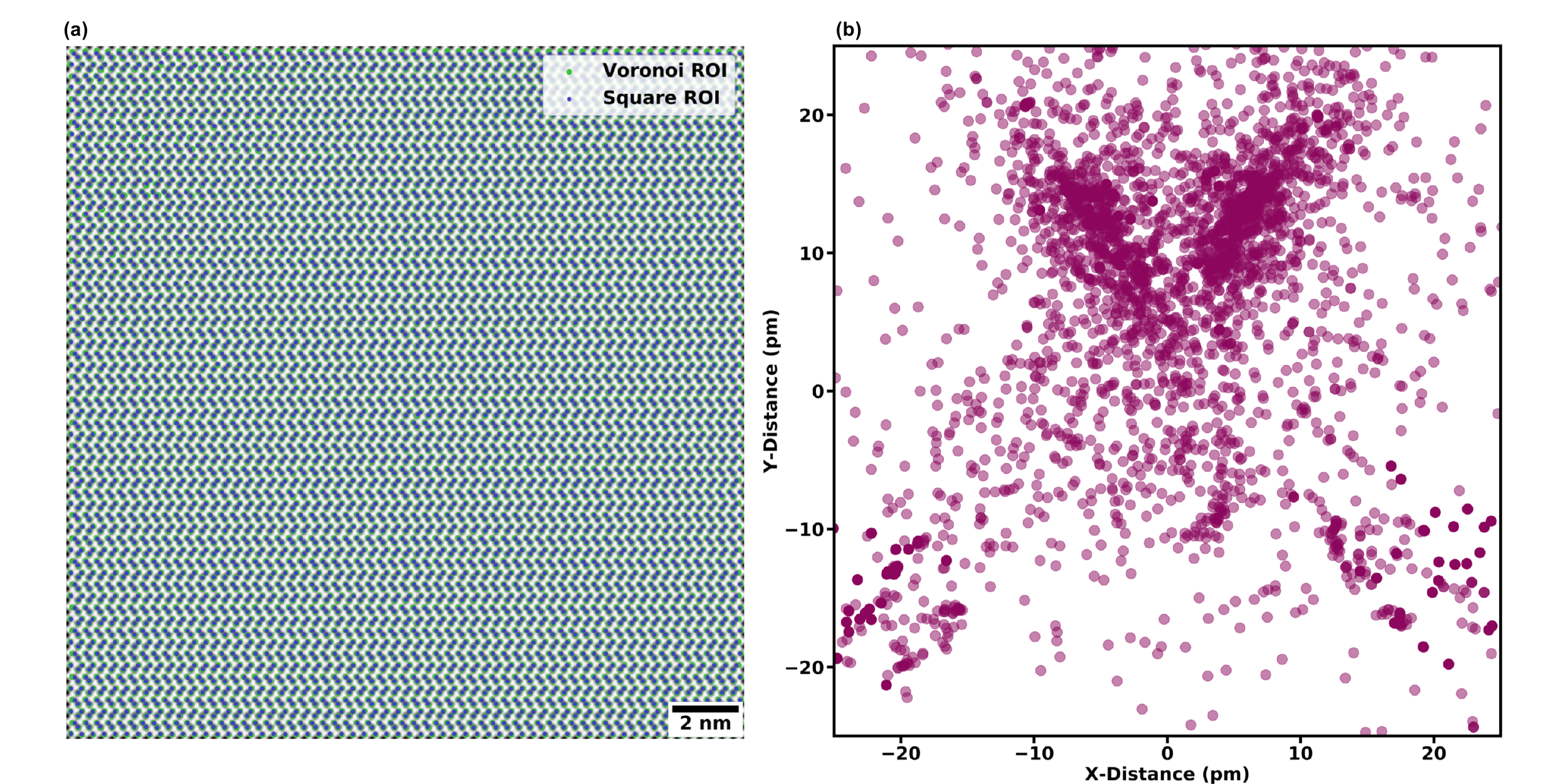}
	\caption{\label{fig:voronoi_ROI} \textbf{Effect of a Voronoi ROI. (a)} Simulated BF-STEM image of \ce{LiNbO3} with the green dots referring to the MPFit calculated positions when the region of interest is a Voronoi region around the intensity minima, while the blue dots refer to refined positions calculated from the standard \textbf{mpfit} algorithm. \textbf{(b)}  Distance between the two results in picometers.}
\end{figure*}

One other region of interest method was tried, apart from the standard box centered on the intensity minima -- a ROI based on the Voronoi region around the intensity minima. A Voronoi region for a point is defined as the section of the image, for which the Cartesian distance to the point is the lowest compared to all other points in the image. However, we observed that the standard ROI actually gave better results rather than the Voronoi tesselation. This is most clearly visible in \autoref{fig:voronoi_ROI}(b) where the distances between the two methods is scattered over even 20pm, while the standard \textbf{mpfit} results are clustered less than 5pm away from the known atom positions as could be observed in \autoref{fig:statstem}(b). The standard deviation $\left( \sigma \right)$ between the positions calculated with the two ROI techniques is 10.93pm, much higher than the standard deviation between the simulated peak positions and the standard \textbf{mpfit} technique.
\end{backmatter}

\bibliographystyle{bmc-mathphys} 
\bibliography{manuscript}

\end{document}